\documentclass[aps,prb,twocolumn,floatfix]{revtex4}
\usepackage{amsmath}
\usepackage{amssymb}
\usepackage{graphicx}
\usepackage{dcolumn}
\usepackage{bm}
\usepackage{wasysym}
\usepackage{color}
\begin{document}

\title{Reduced density-matrix functional theory in quantum Hall systems}

\author{E. T\"ol\"o}
\author{A. Harju} 

\affiliation{Helsinki Institute of Physics and Department of Applied Physics, 
Aalto University, FI-02150 Espoo, Finland}

\begin{abstract}
We apply reduced density-matrix functional theory to the parabolically confined quantum Hall droplet in the spin-frozen strong magnetic field regime. One-body reduced density matrix functional method performs remarkably well in obtaining ground states, energies, and observables derivable from the one-body reduced density matrix for a wide range of system sizes. At the strongly correlated regime, the results go well beyond what can be obtained with the density functional theory. However, some of the detailed properties of the system, such as the edge Green's function, are not produced correctly unless we use the much heavier two-body reduced density matrix method.
\end{abstract}

\maketitle

\section{Introduction}

Quantum Hall fluids occur at low temperature in clean two-dimensional electron systems 
exposed to a perpendicular magnetic field.\cite{klitzing,tsui} Different phases are characterized by the 
number $\nu$ that tells the ratio of the number of electrons to the number of single-particle 
states in a highly degenerate Landau level (the number of flux quanta piercing the sample).
Near certain fractional filling factors $\nu$, the interactions between electrons induce 
an energy gap and lead to a ground state with topological order manifest in exotic 
quasiparticles and non-Fermi liquid edge modes.\cite{laughlin1,laughlin2,halperin1,halperin2,haldane,jain,wen1,wen2,nayak}

Straight from the outset, numerical simulations have been an indispensable guide in development of the theory.\cite{laughlin2}
While majority of the numerics are exact diagonalization studies only viable with small electron numbers, 
variational Monte Carlo\cite{harju} and density functional theory\cite{heinonen,reimann} 
(DFT) have also been applied to larger systems. 
For example, accurate wave functions incorporating the complex non-perturbative effects of electron interactions
have been theoretically devised and later backed up by the high overlap with exact numerical results for small systems.
Due to such developements, reason behind the energy gap of simplest of the many fractions is 
now well understood within the 
framework of composite-fermion theory that allows for explicit construction of the many-body wave function
and calculation of topological quantum numbers.\cite{jain}
However, in going beoynd the composite-fermion theory to more complex phases, the Monte Carlo method is 
crippled since a trustworthy trial wave function for the phases we would be interested to know more about is not known.
In addition, owing to the strong correlations, the DFT is inaccurate at, for example, 
the paradigm fractional quantum Hall state at filling fraction $\nu=1/3$ where the vortices
supposed to form a bound state with the electrons localize at fixed positions instead.\cite{saarikoski2}
Consequently, the exact diagonalization is frequently the only viable alternative, leaving the large electron
numbers beyond the reach of direct calculations, though the density matrix renormalization group method has brought
some progress making a bit larger systems computationally feasible.\cite{shibata,feiguin}

During the past 10 years, reduced density matrix functional theory has been revived and applied successfully in 
the chemical physics community.\cite{muller,ziesche,buijse} The method is known to handle e.g. molecular 
dissociation limits better than standard DFT,\cite{baerends,gritsenko} and it has recently been applied to homogeneous electron 
gas.\cite{lathiotakis1,lathiotakis2} 
In this manuscript, we aim to probe the potential of the 
one-body reduced density-matrix functional theory (1-RDMFT)  in a fractional quantum Hall system, specifically 
a parabolically confined quantum dot in the spin-frozen strong magnetic field regime.
In contrast to many molecular and atomic systems where the dominant occupation numbers are typically close to one, 
this is a highly demanding application for any numerical method as fractional quantum Hall states have long-range quantum entanglement
with all the occupation numbers small for example near 1/3 in the $\nu=1/3$ state.

The performance of various functionals in predicting ground states, 
energies, and observables attainable by the reduced density matrix is compared for small system to the exact diagonalization and
Hartree-Fock with and without the Brillouin-Wigner perturbation theory.
For larger systems with tens of particles, the comparison is done utilizing the accurate Laughlin wave function\cite{laughlin2} for filling 
fraction $\nu=1/3$ state and Monte Carlo methods. Energies are produced quite well in all systems at the strong-correlation 
regime $\nu\ll1$. Even the bulk densities appear reasonably good and reproduce the predicted edge stripe phase.\cite{tsiper}
However, we are still dealing with an approximate method that has its limitations.
Detailed properties of the edge of the electron droplet, such as the edge tunneling exponent,\cite{chang,grayson} are not produced correctly 
by the present functionals.  

For this reason, we also perform a small comparison with the heavier two-body reduced density 
matrix functional theory\cite{garrod} (2-RDMFT) (see the closely related work in Ref.~\onlinecite{rothman}).
Including the exact electron interaction 
by the two-body reduced density-matrix functional appears to facilitate a more accurate description of the edge, however with a 
computational cost in large systems beyond the reach of present day computers.

The rest of the manuscript is organized as follows. In  the next section, we briefly introduce the idea of reduced density matrix
functional methods. In Section~\ref{model}, we describe the model system and derive 
an exact formula for the energy contribution due to
one-body operators present in our Hamiltonian such that only the interaction energy remains to be solved. 
Section \ref{rdmft} and B present the details of our 1-RDMFT and 2-RDMFT implementations, respectively. 
The 1-RDMFT results are divided according to the small or large size of the system into Section \ref{res1} and B, followed 
by the 2-RDMFT calculation in Section \ref{res3}. Conclusions and future prospects are found in Section \ref{summary}.

\section{Reduced density matrix functional theories}
The 1-RDMFT is based on the Gilbert's theorem, which guarantees that the ground-state expectation value of 
any observable is a unique functional of the 1-RDM $\gamma$. \cite{gilbert}
It follows that the ground state energy can be written as
\begin{equation}
\label{frho}
F[\gamma]=\int_{\mathbb{R}^{2d}} \mathrm{d}\mathbf{r}\mathrm{d}\mathbf{r'}\delta(\mathbf{r}-\mathbf{r'})
(T(\mathbf{r})+U(\mathbf{r}))\gamma(\mathbf{r},\mathbf{r'})+V_{\rm ee}[\gamma]
\end{equation}
where $T$ and $U$ are the standard operators for the kinetic energy and an external potential while the functional
for the interaction energy $V_{\rm ee}[\gamma]$ is unknown. It is simple to show that this functional yields 
the exact ground state energy for the exact 1-RDM 
\begin{equation}
\label{exact1rdm}
\begin{split}
\gamma(\mathbf{r},\mathbf{r}')=&N\int\Psi^*(\mathbf{r},\mathbf{r}_2,\mathbf{r}_3,\ldots,\mathbf{r}_N)\\
&\times\Psi(\mathbf{r'},\mathbf{r}_2,\mathbf{r}_3,\ldots,\mathbf{r}_N)\mathrm{d}\mathbf{r}_2\mathrm{d}\mathbf{r}_3\ldots\mathrm{d}\mathbf{r}_N
\end{split}
\end{equation}
if $V_{\rm ee}$ is replaced by half the Coulomb energy of the exact pair-density 
\begin{equation}
\label{exa}
\begin{split}
E_{ee}=&\frac{e^2}{2\epsilon}\int_{\mathbb{R}^{2d}}\mathrm{d}\mathbf{r}\mathrm{d}\mathbf{r'}
\frac{\rho_2(\mathbf{r},\mathbf{r'})}{|\mathbf{r}-\mathbf{r'}|} \ ,\\
\rho_2(\mathbf{r},\mathbf{r}')=&N(N-1)\int\Psi^*(\mathbf{r},\mathbf{r}',\mathbf{r}_3,\ldots,\mathbf{r}_N)\\
&\times\Psi(\mathbf{r},\mathbf{r}',\mathbf{r}_3,\ldots,\mathbf{r}_N)\mathrm{d}\mathbf{r}_3\ldots\mathrm{d}\mathbf{r}_N\ .
\end{split}
\end{equation}
The 2-RDMFT minimizes the resulting exact functional $F$ subject to a subset of complete $N$-representability conditions,
known as the 2-representability conditions, and other possible constraints due to additional symmetries (see Sec.~\ref{2rdm}). On the other hand, 
the crux of the 1-RDMFT is to approximate the pair-density by a functional of the 1-RDM.\cite{herbert} The customary way to do the 
approximation, which we will also employ in this paper, is to replace the pair-density above by
\begin{equation}
\label{apd}
\rho(\mathbf{r})\rho(\mathbf{r'})-\sum_{i,j}f(n_i,n_j)\phi_i^*(\mathbf{r})\phi_j^*(\mathbf{r'})\phi_j(\mathbf{r})\phi_i(\mathbf{r'})
\end{equation}
where $\rho(\mathbf{r})=\gamma(\mathbf{r},\mathbf{r})$ is the density at $\mathbf{r}$, $\phi_i$ are the natural orbitals (eigenvectors of $\gamma$), and $f$ is a function solely of 
the natural occupation numbers $n_i\in[0,1]$ (eigenvalues of $\gamma$).
The functional for the interaction energy in Eq.~(\ref{frho}) then reads
\begin{equation}
\begin{split}
V_{\rm ee}[\gamma]=&\frac{e^2}{2\epsilon}\bigg[\int_{\mathbb{R}^{2d}} \mathrm{d}\mathbf{r}\mathrm{d}\mathbf{r'}\frac{\rho(\mathbf{r})\rho(\mathbf{r'})}{|\mathbf{r}-\mathbf{r'}|}-\sum_{i,j}f(n_i,n_j) \\
&\times\int_{\mathbb{R}^{2d}} \mathrm{d}\mathbf{r}\mathrm{d}\mathbf{r'}
\frac{\phi_i^*(\mathbf{r})\phi_j^*(\mathbf{r'})\phi_j(\mathbf{r})\phi_i(\mathbf{r'})}{|\mathbf{r}-\mathbf{r'}|}\bigg]\ .\\
\label{vee}
\end{split}
\end{equation}
The form of $f$ could vary in different type of systems, whereas those used in this study are 
enlisted in Table~\ref{functions} in Sec.~\ref{rdmft}.
In analogy with the density functional theory, the first term is referred to as the Hartree term while
the latter is the exchange-correlation term. However, compared to the DFT, 1-RDMFT has a couple of advantages.
Firstly, the method obtains not only the density but the whole 1-RDM so that the ground-state expectation value 
of any one-body observable can be readily computed. Thus, for example kinetic and interaction energies can be 
readily separated and Green's functions calculated. Secondly, although both methods are in principle exact for
an exact functional, due to the variable $\gamma$ (vs. $\rho$), it is easier to develop a good 1-RDM functional 
than a good density functional. This is the reason why the DFT does not work in the strongly correlated regime
where the proper treatment of electronic correlations is important. However, there are fresh ideas of how to
treat strongly correlated electrons with DFT. \cite{paola2,paola}

\section{Quantum dot model}
\label{model}

The quantum Hall droplet is modeled by the two-dimensional effective-mass Hamiltonian 
\begin{equation}
\label{ham}
H=\sum^N_{i=1}\left[
 \frac{\left({\bf p}_i+\frac{e}{c} {\bf A}_i \right)^2}{2 m^*}
+\frac{m^*\omega_0^2r^2}{2} \right] +  \sum_{i<j}
\frac{e^2}{\epsilon r_{ij}}\ ,
\end{equation}
where $N$ is the number of electrons, ${\bf A}$ is the planar vector potential
of the homogeneous  magnetic field ${\bf B}$ perpendicular to the sample plane,
and the energy scale of the external confinement potential $\hbar\omega_0$ is typically a few meVs.  
The material parameters are the effective mass of the electrons $m^*=0.067m_e$
and the dielectric constant of the GaAs semiconductor medium $\epsilon=12.7$. 
Coulomb interactions tend to spontaneously polarize the electron spins in an effect known as quantum Hall ferromagnetism.
For this reason, the relatively weak Zeeman term has been omitted and the electrons are assumed spin-polarized.
From here on, we
use oscillator units so that the energies are expressed in units of $\hbar\omega$
and lengths in units of $\sqrt{\hbar/m^*\omega}$ where $\omega^2=\omega_0^2+(\omega_c/2)^2$ with
cyclotron frequency $\omega_c=eB/m^*c$.

For $N=1$, the energy states are written as
\begin{equation}
\label{wff}
\psi_n^m(z)=\sqrt{\frac{n!}{\pi(n+m)!}}z^mL_n^m(z\bar{z})e^{-z\bar{z}/2} ,\  m\geqslant-n ,\  n\geqslant0
\end{equation}
where $z=x+iy$ and $L_n^m$ are the generalized Laguerre polynomials. The corresponding eigenvalues are given by 
\begin{equation}
\label{ene}
E_n^m=2n+1+\left(1-\frac{\hbar\omega_c}{2}\right)m\ .
\end{equation}
We take on interest in developing a computational method for the fractional quantum Hall states 
at the strong magnetic field regime, so we may assume that  $\omega_0\ll\omega_c$. 
Then $\hbar\omega_c/2$ is close to unity, such that the term in
parentheses in Eq.~(\ref{ene}) becomes small and values of quantum number $n$ other than 0 become irrelevant to the 
low energy physics. This is the Landau level projection to the band with $n=0$. It should not be difficult to generalize the
1-RDMFT method to include spin and several Landau levels and study the region $\nu>1$ as well, however, for simplicity we stick to
the spin-polarized lowest Landau level $\nu\leqslant1$ in this study.

Note that in our two dimensional model, $m$ in Eq.~(\ref{wff}) is the one and only angular momentum quantum number. We can write the contribution
to the total energy due to terms other than the Coulomb interaction for a system of $N$ electrons in the lowest Landau level $n=0$
in terms of the total angular momentum (quantum number) $M$ exactly as
\begin{equation}
\label{tu}
T+U=\sum_{i=1}^N\left[1+\left(1-\frac{\hbar\omega_c}{2}\right)m_i\right]=N+\left(1-\frac{\hbar\omega_c}{2}\right)M\ .
\end{equation}

As the total angular momentum operator $\hat{M}$ commutes with the total Coulomb interaction energy operator $V_{\rm ee}$, 
to solve the Landau level projected spectrum, the remaining task is to find the common
eigenstates of the Landau level projected $V_{\rm ee}$ and $\hat{M}$.

For $N$ and $M$ small enough, these are solved exactly with the configuration interaction method (exact diagonalization)
since the number of possible single-particle states is finite. 
We may go to a bit larger $N$ and $M$ by taking 
interest in only the ground state and finding it by the Lanczos algorithm (cf. Refs. \onlinecite{tolo,tolo2}).
However, the exponential growth of the many-body basis limits the particle number to around 10 depending on $M$,
and some kind of an approximate calculation method becomes necessary. The Kohn-Sham DFT is 
still a good method for the weakly correlated regime\cite{saarikoski1} but for larger $M$ the natural occupations tend far from 1 
and the DFT no longer gives us good results. 
Absence of a reliable trial wave function for generic $M$ makes the implementation of variational Monte Carlo rather uncertain.\cite{harju,siljamaki}
In this paper we are going to see, if the reduced density matrix functional theory can make itself useful. The
expectation is that it will work better than DFT at least when the eigenvalues of the density matrix are small meaning $\nu\ll1$.
With several Landau levels, this would correspond to the situation where the highest occupied Landau level has low filling fraction.

\section{Computational methods}
\subsection{1-RDMFT}
\label{rdmft}

Due to the exact formula for the one-body operators' energy contribution (Eq.~(\ref{tu})), the energy functional of Eq.~(\ref{frho}) simplifies to
\begin{equation}
F[\gamma]=N+\left(1-\frac{\hbar\omega_c}{2}\right)M+V_{\rm ee}[\gamma]\ ,
\end{equation}
with the constraints that the particle number and the  total angular momentum are $N$ and $M$, respectively.
Moreover, the natural orbitals for energy state $|\Psi\rangle$ in the lowest Landau level are directly
the single-particle energy states since the angular momentum conservation yields a diagonal density matrix
in this basis
\begin{equation}
\gamma_{mm'}=\langle\Psi|a_m^{\dagger}a_{m'}|\Psi\rangle=\delta_{mm'}n_m\ ,
\end{equation}
where $a_m$ and its adjoint annihilate and create a particle with quantum numbers $n=0$ and $m$ (see Eq.~(\ref{wff})).
As $N-1$ particles have at least a total angular momentum $(N-1)(N-2)/2$, the maximum single-particle
angular momentum becomes $k=M-(N-1)(N-2)/2$.
Therefore, the minimization of the interaction energy reduces to the constrained minimization of a function whose
variables are $k+1$ occupation numbers
\begin{equation}
V_{\rm ee}(\{n_i\}_{i=0}^k)=\frac{1}{2}\
\sum_{i,j}(n_i n_{j} V_{ijij}-f( n_i,n_{j} )V_{ijji}) 
\end{equation}
where $V_{ijkl}$ are the interaction matrix elements of the lowest Landau level orbitals computed in Ref. \onlinecite{tsiper2}, and 
the constraints are explicitly written as
\begin{equation}
\sum_{m=0}^kn_m=N \textrm { and }\sum_{m=0}^kn_mm=M\ .
\end{equation}
To optimize the occupation numbers, we first express them in terms of variables $\theta_i$ 
such that $n_i=\sin^2\theta_i$. The minimization of $V_{\rm ee}$ with the above two remaining constraints 
is then performed with the interior point or Nelder-Mead algorithm of Mathematica.\cite{mathematica}

A vast number of functions $f( n_i,n_{j} )$ have been proposed in the literature in treatment of simple atoms and molecules, 
and it is not immediately clear, which of them would work well in the current system. Table~\ref{functions} summarizes 
those used in this work for small systems to find out the optimal ones. For large systems, we only use the density-matrix power functional
$f(n_i,n_j)=(n_i,n_j)^{\alpha}$ (P-$\alpha$).\cite{lathiotakis2}
Each form of the off-diagonal $f(n_i,n_{j\neq i})$ corresponds to two
different functions, first of which has diagonal $f(n_i,n_i)=n_i$ while the second has $f(n_i,n_i)=n_i^2$.
Since the natural orbitals are orthonormal, integration of the approximate pair-density in Eq.~(\ref{apd}) 
over the coordinates yields the proper normalization $N(N-1)$ for $f(n_i,n_i)=n_i$. The form $n_i^2$, on the
other hand, is justified as it cancels the self-interaction terms $V_{iiii}$ arising from the Hartree term.

A few words about different forms of the off-diagonal terms are in order (see last column in Table~\ref{functions}).
The form $n_i^{\alpha}n_j^{1-\alpha}$ for $f$ was first introduced by M\"uller, who found $\alpha=1/2$ (MU) to
be the optimal value.\cite{muller} Goedecker and Umrigar (GU) used the modification to the diagonal that removes 
the self-interaction terms.\cite{goedecker} Much similar to the density-matrix power functional,
we generalize these to MU-$\alpha$ and GU-$\alpha$ where the square root is replaced by an arbitrary power, presumably lying between half and one. 
Alternatively, it is 
often physically motivated to reduce 
the overcorrelation of MU by switching the sign of the off-diagonal terms between weakly occupied natural orbitals $W$ (BBC1) 
or additionally banishing the square root for pairs of strongly occupied orbitals $S$ (BBC2). \cite{lathiotakis1,gritsenko}
For simplicity, we define the strongly and weakly occupied orbitals directly from the Hartree-Fock solution where the occupations are 0 or 1.

\begin{table}[htb]
\caption{Functions $f(n_i,n_j)$. $S$ and $W$ refer to the strongly and weakly occupied natural orbitals, respectively.}
\label{functions}
\begin{displaymath}
\begin{array}{r|c|c}
&f(n_i,n_i)&f(n_i,n_{j\neq i})\\
\hline
\begin{array}{r}
\textrm{MU}\\
\textrm{GU}
\end{array}
&
\begin{array}{c}
n_i\\
n_i^2
\end{array}
&
\sqrt{n_in_j}\\
\hline
\begin{array}{r}
\textrm{MU-$\alpha$}\\
\textrm{GU-$\alpha$}\\
\textrm{P-$\alpha$}
\end{array}
&
\begin{array}{c}
n_i\\
n_i^2\\
n_i^{2\alpha}
\end{array}
&
(n_in_j)^{\alpha}\\
\hline
\begin{array}{r}
\textrm{BBC1}\\
\textrm{BBC1S}
\end{array}
&
\begin{array}{c}
n_i\\
n_i^2
\end{array}
&
\left\{
\begin{array}{rl}
-\sqrt{n_in_j}&\ i,j\in W\\
\sqrt{n_in_j}&\ i,j\notin W\\
\end{array}\right.\\
\hline
\begin{array}{r}
\textrm{BBC2}\\
\textrm{BBC2S}
\end{array}
&
\begin{array}{c}
n_i\\
n_i^2
\end{array}
&
\left\{
\begin{array}{rl}
-\sqrt{n_in_j}&\ i,j\in W\\
n_in_j&\ i,j\in S\\
\sqrt{n_in_j}&\textrm{else}\\
\end{array}\right.\\
\end{array}
\end{displaymath}
\end{table}

Finally, we would like to point out an issue with the physicality of the
obtained solution, which is generally more of a problem in the higher order 
RDM methods. Specifically, Coleman has shown that necessary and sufficient
condition for the 1-RDM to be $N$-representable, meaning that
there exists $|\Psi\rangle$ in the Hilbert space of the system such that 
$\gamma_{ij}=\langle\Psi|a_i^{\dagger}a_j|\Psi\rangle$, is that its eigenvalues
are between 0 and 1 and their sum is $N$.\cite{coleman} However, if we additionally demand
a symmetry, it may be that the obtained solution is not representable in the
symmetry restricted part of the Hilbert space. To be explicit, if in our model
we demand total angular momentum of a two electron system to be 2, the symmetry
restricted Hilbert space of states with $m=0,1$, and 2 consists only of one state with occupations $(1,0,1)$ while 
1-RDM method could give us unphysical occupations $(0.5,1,0.5)$ both having the same
particle number and angular momentum ($0.5\times0+1\times1+0.5\times2=2$). While this could be avoided by imposing
additional constraints, we refrain from doing it as that would be exponentially unfeasible 
for larger systems. Additionally, it is plausible that the unphysical solutions have less weight
when the size of the physical Hilbert space increases.

\subsection{2-RDMFT}
\label{2rdm}
In analogy with the 1-RDMFT, we now minimize a functional of the 2-RDM
\begin{equation}
\label{gijkl}
\Gamma^{ij}_{kl}=\langle\Psi|a_i^{\dagger}a_j^{\dagger}a_la_k|\Psi\rangle\ .
\end{equation}
The marked difference is that we now have an exact functional for the interaction energy 
\begin{equation}
\label{2fun}
V_{\rm ee}(\Gamma)=\frac{1}{2}\
\sum_{i,j,k,l}\Gamma^{ij}_{kl}V_{ijkl}\ ,
\end{equation}
however, with
the cost of large number of additional parameters to optimize with similarly large number of additional constraints. 
Furthermore, the set of constraints to be listed below form a relatively stringent set of necessary
conditions that only in special cases is sufficient for the obtained solution  to be exactly
$N$-representable, meaning a physical $|\Psi\rangle$ to exist such that Eq.~(\ref{gijkl}) holds.

The minimization of the functional (\ref{2fun}) is performed  by forming an augmented Lagrangian function
and minimizing it following the algorithm in Ref.~\onlinecite{mazziotti}. The minimization is performed
with limited memory quasi-Newton algorithm of Mathematica, and the form of the augmented Lagrangian is
\begin{equation}
\label{ala}
L=F[\gamma]-\sum_i\lambda_ic_i+\sum_ic_i^2/{\mu}
\end{equation}
where $\lambda_i$ are the Lagrange multipliers and $\mu>0$ is the augmentation parameter used to enforce 
the convergence of the constraints $c_i=0$.
In the following, we first introduce the subset of applied $N$-representability conditions, and after that, impose
the further constraints due to the fixed total angular momentum, $M$-representability.

\subsubsection{$N$-representability}
The trace condition
\begin{equation}
\label{ntr}
\sum_{i<j}\Gamma^{ij}_{ij}=\frac{N(N-1)}{2}
\end{equation}
is used to fix the particle number to $N$.

Positivity conditions form the major part of the $N$-representability constraints. Consider an operator
of the form $A=\sum_{i_1i_2\ldots i_k} t_{i_1i_2\ldots i_k}a_{i_1}a_{i_2}\ldots a_{i_k}$.
Since $\langle\Psi|A^{\dagger}A|\Psi\rangle\geqslant0$, it follows that 
\begin{equation}
\sum_{\substack{i_1i_2\ldots i_k\\j_1j_2\ldots j_k}}t^*_{i_1i_2\ldots i_k}t_{j_1j_2\ldots j_k}\Gamma^{i_1i_2\ldots i_k}_{j_1j_2\ldots j_k}\geqslant0\ .
\end{equation}
For $k=2$ we obtain the 2-positivity condition for the 2-RDM
\begin{equation}
\sum_{\substack{ij\\kl}}t^*_{ij}t_{kl}\Gamma^{ij}_{kl}\geqslant0\ .
\end{equation}
By a different choice of $A$, positivity conditions of the exact same form can be derived for the two other
representations of the 2-RDM
\begin{equation}
\begin{split}
Q^{ij}_{kl}=&\langle\Psi|a_ia_ja_l^{\dagger}a_k^{\dagger}|\Psi\rangle\textrm{ and}\\
G^{ij}_{kl}=&\langle\Psi|a_i^{\dagger}a_ja_l^{\dagger}a_k|\Psi\rangle\ .
\end{split}
\end{equation}
While the representations $\Gamma$, $Q$, and $G$ are all equivalent as they are related by the fermionic anticommutation rules,
the positivity conditions are inequivalent and must be taken into account simultaneously. We use the 
antisymmetric basis $|(ij)\rangle=(|ij\rangle-|ji\rangle)/2$ for $\Gamma$ and $Q$ matrices since 
$\Gamma^{ij}_{kl}=-\Gamma^{ji}_{kl}=-\Gamma^{ij}_{lk}$ and $Q^{ij}_{kl}=-Q^{ji}_{kl}=-Q^{ij}_{lk}$. 
Furthermore following Ref.~\onlinecite{mazziotti}, since all the three matrices are real and symmetric under $ij\leftrightarrow kl$, the
positive-definite condition can be accounted for simply by writing the matrices as square of symmetric matrices 
$\Gamma=R^2$, $Q=S^2$ , and $G=T^2$ (meaning $\Gamma^{(ij)}_{(kl)}=\sum_{(pq)}R^{(ij)}_{(pq)}R^{(pq)}_{(kl)}$
etc.) and optimizing the upper diagonal of matrices $R$, $S$, and $T$. The linear relations linking $\Gamma$ and $Q$ and
$\Gamma$ and $G$ become the relevant constraint equations
\begin{equation}
\label{qg}
\begin{split}
Q^{ij}_{kl}=&\Gamma^{ij}_{kl}-\delta_{ik}\gamma_{jl}-\delta_{jl}\gamma_{ik}+\delta_{il}\gamma_{jk}+\\
&\delta_{jk}\gamma_{il}+\delta_{ik}\delta_{jl}-\delta_{il}\delta_{jk}\ ,\\
G^{ij}_{kl}=&\delta_{jl}\gamma_{ik}-\Gamma^{il}_{kj}\ ,\\
\end{split}
\end{equation}\\
where the 1-RDM $\gamma_{ij}$ may be obtained through
\begin{equation}
\label{rdmrel}
\sum_{k}G^{ij}_{kk}=N\gamma_{ij} \textrm{ or } \sum_{k}\Gamma^{ik}_{jk}=(N-1)\gamma_{ij}.
\end{equation}

\subsubsection{$M$-representability}
Since $\sum_im_ia_i^{\dagger}a_i|\Psi\rangle=M|\Psi\rangle$, we can form one independent 
nontrivial equation involving 2-RDM (equivalent to the contracted Schr\"odinger equation in Ref. \onlinecite{rothman}) 

\begin{equation}
\label{mc1}
\sum_{k}m_kG^{ij}_{kk}=M\gamma_{ij}\ .
\end{equation}
The trace of this is already fixed by Eqs.~(\ref{ntr}) and (\ref{rdmrel})
\begin{equation}
\sum_{ij}m_jG^{ii}_{jj}=MN\ .
\end{equation}
Since in our system the angular momentum quantum numbers $m_j=j$ are always non-negative, 
the maximum angular momentum for
a pair of electrons to have is $k_2=M-(N-2)(N-3)/2$ where the subtracted term is the
minimum angular momentum of $N-2$ electrons.
Additionally, some of the matrix elements of the 2-RDM are zero because states with different
total angular  momentum are orthogonal. The independent constraints due to these considerations read
\begin{equation}
\label{mc2}
\Gamma^{ij}_{kl}=0\textrm{ , }i+j\neq k+l\textrm{ or }i+j>k_2\ ,
\end{equation}
while Eq.~(\ref{qg}) communicates them to $Q$ and $G$. Though we can just drop the corresponding
terms from our equations, we still need to take into account the ensuing constraints on the actual 
variables $R$,$S$, and $T$.

\subsection{Monte Carlo}

The calculation of the natural orbital occupations is relatively simple using the Monte Carlo technique. Unlike in the
previous Monte Carlo study in Ref. \onlinecite{kent}, we know from the start the natural orbitals that diagonalize the
density matrix, and thus we only need to calculate the occupations.

Starting from the definition (\ref{exact1rdm}), the orthogonality of the natural orbitals, and the expansion of 1-RDM using the natural orbitals
\begin{equation}
\gamma(\mathbf{r},\mathbf{r}')=\sum_mn_m\phi_m^*(\mathbf{r})\phi_m(\mathbf{r}')\ ,
\end{equation}
the occupations are integrated as
\begin{equation}
\begin{split}
n_m =& N \int \Psi^*(\mathbf{r}_1,\mathbf{r}_2,\dots)
\Psi(\mathbf{r}',\mathbf{r}_2,\dots)\\
&\times\phi_m(\mathbf{r}_1) \phi_m^*(\mathbf{r}') d\mathbf{r}' d\mathbf{r}_1 \dots d\mathbf{r}_N\ .
\end{split}
\end{equation}
This is further reformulated as
\begin{equation}
\begin{split}
n_m =& N \int |\Psi(\mathbf{r}_1,\mathbf{r}_2,\dots)|^2
\frac{\Psi(\mathbf{r}',\mathbf{r}_2,\dots)}{\Psi(\mathbf{r}_1,\mathbf{r}_2,\dots)}
 \\
&\times\phi_m(\mathbf{r}_1) \phi^*_m(\mathbf{r}') d\mathbf{r}' d\mathbf{r}_1 \dots d\mathbf{r}_N\ ,
\end{split}
\end{equation}
which can be symmetrized and rewritten in Monte Carlo expectation value as
$$
n_m= \left\langle \sum_i
\int \frac{\Psi(\mathbf{r}',\mathbf{r}_2,\dots)}{\Psi(\mathbf{r}_1,\mathbf{r}_2,\dots)} \phi_m(\mathbf{r}_i)
\phi^*_m(\mathbf{r}') d\mathbf{r}'
\right\rangle_{\{\mathbf{r}_i\} \in |\Psi|^2} 
$$
where the summation is over the coordinates $\{\mathbf{r}_i\}_{i=1}^N$.
The first strategy for Monte Carlo evaluation of the occupation is to sample
these coordinates from $|\Psi|^2$ and to integrate over
$\mathbf{r}'$ on a grid.\cite{kent}

For a better option in our case, we first rewrite
$
\phi_m(\mathbf{r}_i) \phi^*_m(\mathbf{r}') = |\phi_m(\mathbf{r}')|^2 \frac{\phi_m(\mathbf{r}_i)}{ \phi_m(\mathbf{r}')}
$
and then do a Monte Carlo integration also over $\mathbf{r}'$ as
\begin{equation}
n_m= \left\langle \sum_i
\frac{\Psi(\mathbf{r}',\mathbf{r}_2,\dots)}{\Psi(\mathbf{r}_1,\mathbf{r}_2,\dots)}
\frac{\phi_m(\mathbf{r}_i)}{\phi_m(\mathbf{r}')}
\right\rangle_{\{\mathbf{r}_i\} \in |\Psi|^2,  \ \mathbf{r}' \in
  |\phi_m|^2} 
\end{equation}
where $\{\mathbf{r}_i\}_{i=1}^N$ is again sampled from $|\Psi|^2$ and
$\mathbf{r}'$ from $|\phi_m|^2$. This option can be made more stable by noting
that the natural orbitals have rotation symmetry and
$|\phi_m(\mathbf{r}')|^2$ depends only on the radial coordinate $r'$ and not
on the angle $\theta '$. Now, the radial integral over $r'$ can be done using
Monte Carlo integration and the angular integral by averaging over a uniform
grid $\{\theta'_j\}_{j=1}^{N_{\theta'}}$ as
$$
n_m= \left\langle \frac{1}{N_{\theta'}}\sum_{i, j}
\frac{\Psi(\mathbf{r}',\mathbf{r}_2,\dots)}{\Psi(\mathbf{r}_1,\mathbf{r}_2,\dots)}
\frac{\phi_m({r}_i,\theta_i)}{\phi_m({r}',\theta'_j)}
\right\rangle_{\{\mathbf{r}_i\} \in |\Psi|^2,  \ r' \in
  |\phi_m|^2}  \hspace{-2.3cm} .\hspace{2.3cm}
$$
Notice that $r'$ is generated separately for each $\phi_m$.

\section{Results}
\label{results}
In the following, we will first compare the performance of various 1-RDM functionals in calculating
the interaction energies in different $(N,M)$ sectors in exactly solvable small quantum Hall droplets. 
This analysis is deepened by analyzing the occupation numbers obtained for the ground states. Once
we have established that we have a decent functional, we proceed to test its performance in larger systems.
Since no exact results are available for larger systems at the strongly correlated regime $\nu\ll1$, we employ
the next best gripping handle, which is the Laughlin's variational wave function for filling fraction $\nu=1/3$.
Moreover, we compare the energies and occupations obtained with density-matrix power functionals (P-$\alpha$) to the results
extracted from the Laughlin's wave function with Monte Carlo techniques. From the occupation
numbers we calculate the edge Green's function $G_{\rm E}$, which gives information about the topological order
of different quantum Hall phases. It is interesting to see if the newly applied methods (1-RDMFT and 2-RDMFT) reproduce 
the correct decay properties of $G_{\rm E}$.

\subsection{Towards a good 1-RDM functional in small quantum Hall droplets}
\label{res1}
The energy states of the quantum dot may be written as
\begin{equation}
E^n_{MN}=N\hbar\omega+(\hbar\omega-\hbar\omega_c/2)M+V_{\rm ee}^n(M,N)\ ,
\end{equation}
where $V^n_{ee}(M,N)$ is the $n$th eigenvalue of the total interaction energy in sector $(M,N)$, whose
dependence on the physical parameters is scaling by a factor $e^2/\epsilon l$.
Because of the second term, it follows that all possible ground states for $N$ particles are found at the 
intersections of the $M$-$V_{\rm ee}$-curve and its convex lower envelope. 

\begin{figure*}[htbp]
\begin{center}
\includegraphics[width=2\columnwidth]{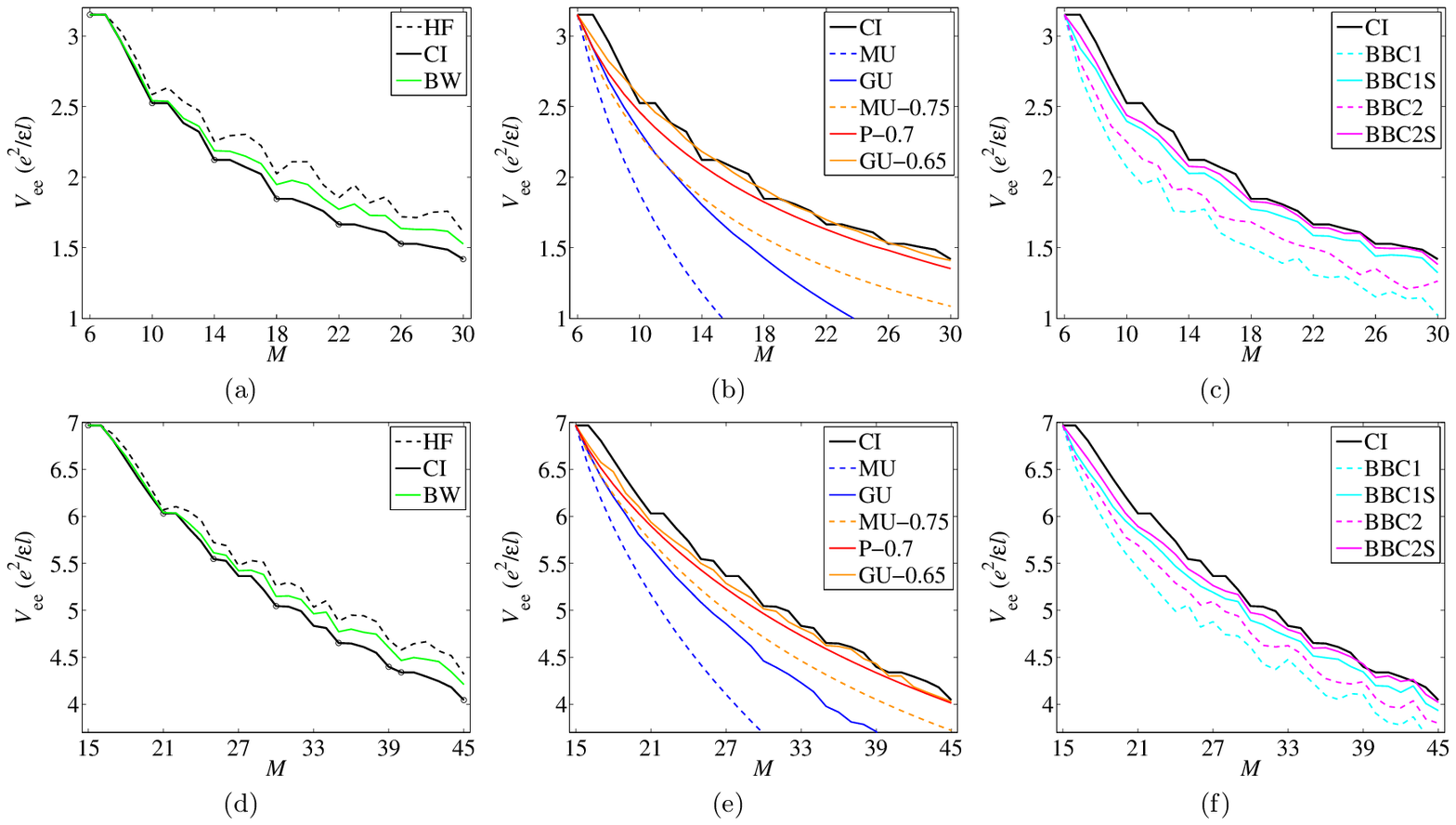}
\caption{(Color online) The minimum interaction energy $V_\mathrm{ee}$ 
at each angular momentum $M$ for 4 (a-c) and 6 electrons (d-f). The exact diagonalization ground states 
detected by the convex envelope are marked with circles in (a) and (d).}
\label{vil}
\includegraphics[width=2\columnwidth]{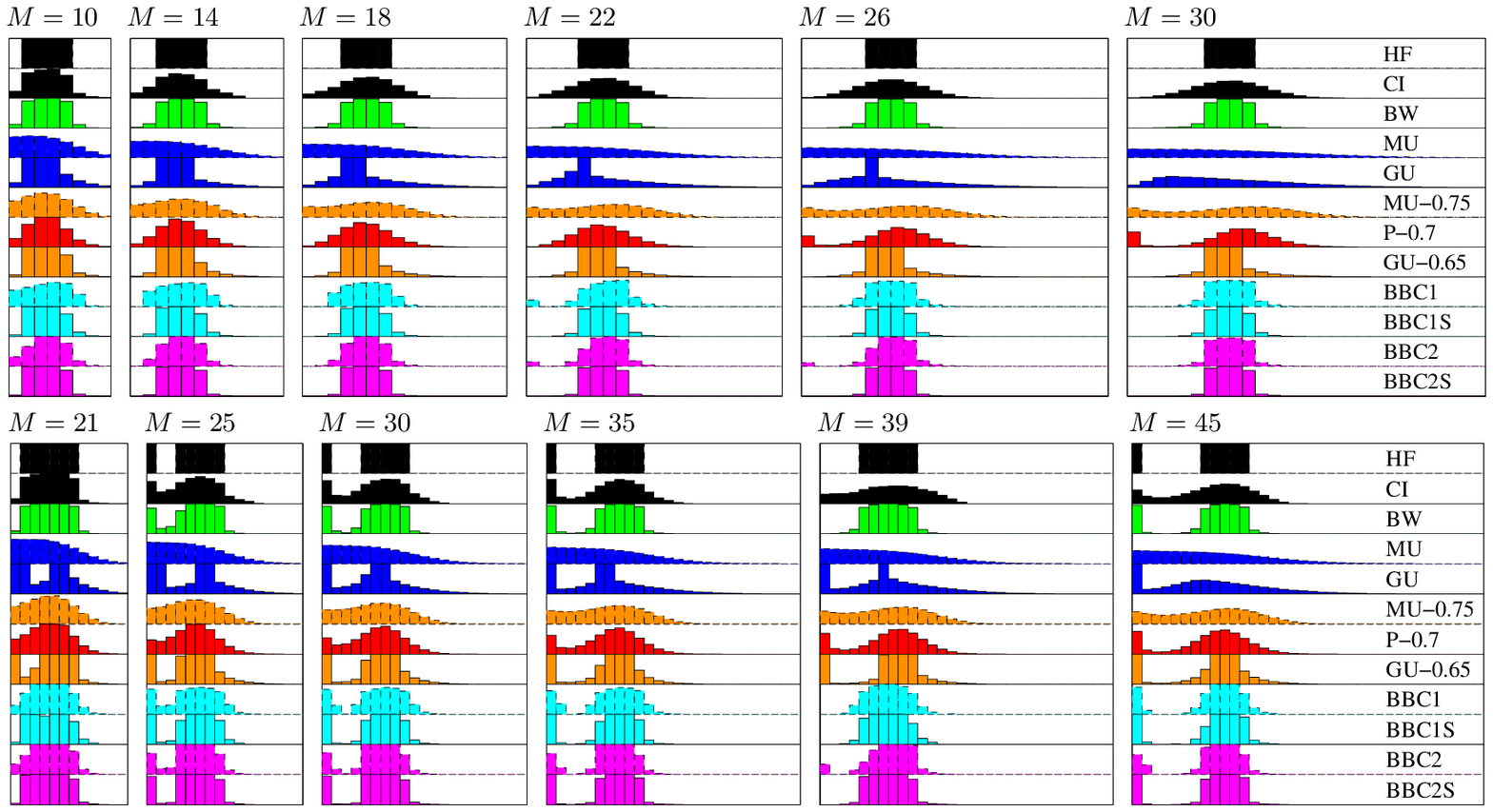}
\caption{(Color online) The occupation numbers for selection of  the 4-electron (upper panel) and 6-electron (lower panel) ground states 
indicated in Fig.~\ref{vil} (a) and (d), respectively. A bar that fills the assigned height corresponds to occupation number 1.
Occupations are ordered according to the increasing single-particle angular momentum $m$ of the lowest Landau level orbitals
starting with 0 at the left. Because most likely location of electron at orbital $m$ is at distance $r=\sqrt{m}$ from the center,
one can think the set of orbitals as a radially discretized disk.}
\label{oc}
\end{center}
\end{figure*}

Figure \ref{vil}(a) shows the $M$-$V_{\rm ee}$-curves for 4 electrons computed with the configuration interaction (CI), Hartree-Fock method (HF), and the Brillouin-Wigner 2nd order perturbation theory (BW) to the HF state.
The exact diagonalization (CI) ground states, detected by the convex envelope, are marked by circles. While both HF and BW predict 
the correct ground states, the perturbation theory leads to a significant improvement to the HF energy. The 2nd order perturbation
theory is very accurate and close to the CI result for $M<14$, and it is roughly half-way between HF and CI energy for $M>14$. 

Physically the cusp structure seen in the results follows from the energetic advantage of a configuration, in which
the 4 electrons are located at vertices of a square. This configuration has non-zero weight only if the
angular momentum attains a special value such that $N(N-1)/2\equiv\mathrm{mod}(M,N)$. The difference between
subsequent magic angular momentum states is the number of vortices found at the center of the system. As magnetic field
is increased, vortices that carry quantized angular momentum, and in a sense quantized magnetic flux, emerge at
the center.
As the particle number is taken very large, the number of cusps increases while all cusps no 
longer correspond to a ground-state at certain system parameters. Instead, a few of them are more special than others
forming the incompressible vacuum state of certain fractional quantum Hall phase 
as the remaining cusps are
related to the quasiparticle/vortex excitations that are eventually responsible for the finite extension of the quantum Hall plateaus.

The corresponding energies obtained with 1-RDMFT are shown in Figs.~\ref{vil}(b) and (c). At this point the choice of 
parameters $\alpha=0.75$, $\alpha=0.7$, and $\alpha=0.65$ for MU-$\alpha$, P-$\alpha$, and GU-$\alpha$ is an educated guess, 
whereas the effect of the parameter $\alpha$ becomes apparent in the next section (basically, it tunes the strength of electron
correlations).
Typical to 1-RDMFT calculations in general,
the energies are below
the CI result. The dashed lines systematically lie below the solid lines of the same tone, due to the self-energy cancellation
present in the latter.
Basic MU and GU functionals clearly overestimate the correlation energy and behave even qualitatively wrong as they fall too fast
with increasing $M$.
For the rest of the functionals, $V_{\rm ee}$ seems to decline at about the correct rate as a function of $M$.
However, a nice cusp structure is only seen with the BBC1S and BBC2S functionals, which inherit the cusps from the HF
state used in the selection of the strongly and weakly occupied orbitals.
Overall the energetically best of these 1-RDM functionals (P-0.7, GU-0.65, BBC1S, and BBC2S)
perform better than the 2nd order perturbation theory when $M>14$, although only BBC2S produces the correct 
ground state structure.

The equivalent curves for 6 electrons are shown in Figs.~\ref{vil} (d)-(f). As the 6-electron results are quantitatively 
like the 4-electron results, the performance of different functionals seems to be rather insensitive to the particle number. 
For six electrons, the cusps should occur at $N(N-1)/2\equiv\mathrm{mod}(M,N)$ or $N(N-1)/2\equiv\mod(M,N-1)$ corresponding to
a hexagonal configuration or a pentagonal configuration with one electron at the center.
All the cusps are not correctly reproduced with any 1-RDM functional, though BBC2S result follows
the cusp structure quite well at $M>27$.

The quantum Hall droplet model has the property that an increase in $M$ increases the area of the droplet and also the
electronic correlations quantified in reduction of the occupation numbers of the natural Landau level orbitals. 
The average occupation number of the relevant orbitals is close to the corresponding macroscopic quantum Hall 
filling fraction $\nu$ and becomes exact as $N$ is taken to infinity. For example, the $\nu=1$ state occurs at
the minimum angular momentum $M=N(N-1)/2$ where the first $N$ orbitals have occupation 1. 
Second example is the Laughlin's wave function\cite{laughlin2} for filling fraction $\nu=1/3$ 
\begin{equation}
\label{lawf}
\Psi^{1/3}_{\textrm{L}}(\{z_i\})=\prod_{i<j}(z_i-z_j)^3e^{-\frac{1}{2}\sum_i z_i\bar{z}_i}\ .
\end{equation}
It has angular momentum $M=3N(N-1)/2$ and $3(N-1)+1$ nonempty orbitals such that on average the fraction $\nu=N/(3(N-1)+1)\approx1/3$ is filled.
The Laughlin state has about 0.98 overlap with the exact ground state for 4 and 6 electrons,
and it is the lowest angular momentum zero-energy state of the short-range model interaction\cite{trugman}
\begin{equation}
V_1(z_{ij})=\partial_{z_i}\partial_{\bar{z}_i}\delta(z_i-z_j)+i\leftrightarrow j\ .
\label{pseudopotential}
\end{equation}
Note, however that while for four electrons the highly correlated Laughlin state occurs at $M=18$ near the center 
of the $M$-window in the $M$-$V_{\rm ee}$-curves, the corresponding angular momentum for six electrons is $M=45$, and 
it is the highest $M$ included in the corresponding figures. Overall, it seems that the perturbation theory works
energetically well near $\nu=1$ where the correlations are weak while the 1-RDM functionals perform better at
the strongly correlated regime $\nu\ll 1$, which raises some hope for the 1-RDMFT to prove valuable in quantum
Hall systems.

But how close are the obtained minimizing 1-RDMs actually to the exact results? Recall that the natural orbitals in the lowest
Landau level are fixed and their occupations completely specify the 1-RDM. Figure \ref{oc} shows the 
occupation numbers corresponding to the ground states indicated in Fig.~\ref{vil} (a) and (d). 

Looking
at the first row of occupation numbers for 4 electrons, the next HF ground state is obtained from the previous by adding 
a hole to the center leading to angular momentum increase $N$. The exact CI result below is similar but, in addition,
the correlations spread the occupations at each step. On the third row, the second order perturbation 
theory BW has a small spread of occupations in accordance with the energy curves in Fig.~\ref{vil}(a). 
The 1-RDM functional results on the following 9 rows
are varied in nature. In accord with the poor energies, MU functional leads to a way too large spread of occupation 
and so does the GU, although the latter also pins some occupations to one. The inclination towards pinning is 
due to the self-energy cancellation, since without the cancellation non-pinned occupations lead to negative 
self-energy contribution lowering the total energy ($n_i^2-n_i<0$ for $0<n_i<1$). This is the reason why MU-0.75 is more
spread out than GU-0.65, and BBC occupations are a bit less pinned than BBCS occupations. 
Despite the better energetics, the GU-0.65, BBC1S, and BBC2s occupations seem not to be much better than the second 
order perturbation theory. On the contrary, P-0.7, on the other hand, has both quite good energy and occupations numbers
only slightly less spread than the exact result. Note however, the nonzero first occupation at $M=26$ and $M=30$ in
contrast to the exact result.

For six electrons (Fig.~\ref{vil} (lower)), the occupations are similar. Note the high probability for
one electron to be at the center in some of the HF and CI ground states, correctly reproduced by many of the functionals. 
The six-electron occupation numbers at $M=30$ and $M=35$ can be directly compared to the occupations obtained with DFT in Ref.~\onlinecite{saarikoski2},
and they are found to be a bit similar to our BBC1 or BBC2 results.

On the whole, the P-0.7 power functional seems like a good candidate functional for systems with large number of electrons.
GU-$\alpha$ with $\alpha <0.65$ and MU-$\alpha$ with $\alpha>0.75$ could also work well, however, since the diagonal
part of the P-$\alpha$ functional is somewhat a compromise between these two, we employ the P-$\alpha$ functional
in the remainder of the manuscript.

\subsection{1-RDM at large $N$}

\label{res2}

The results of the previous section suggest that the density matrix power functional (P-$\alpha$) could be a good functional
in quantum Hall systems, 
and thus we apply it to large systems for a few parameters $\alpha$.
Moreover, we concentrate on the $\nu=1/3$ state, whereby close to exact nonperturbative results can be computed with
the Laughlin's variational wave function (Eq.~(\ref{lawf})) using Monte Carlo. It is natural to limit the number of natural 
orbitals to that of the Laughlin wave function $3(N-1)+1$ although the realistic Coulomb ground state would actually 
extend, weakly though, to a few more orbitals. 

\begin{figure*}[htbp]
\begin{center}
\includegraphics[width=2\columnwidth]{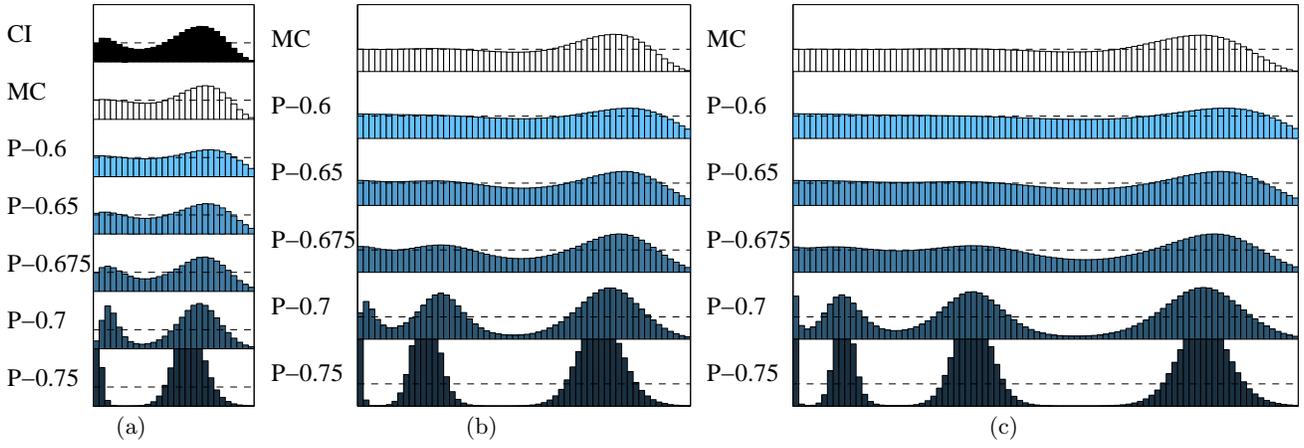}
\caption{(Color online) The occupation numbers at the angular momentum of the $1/3$ Laughlin state for (a) $N=10$, (b) $N=20$, 
and (c) $N=30$ electrons with $3(N-1)+1$ natural orbitals. MC is the exact Laughlin wave function result extracted with Monte Carlo.
The dashed line marks $1/3$ occupation.}
\label{moc}
\end{center}
\end{figure*}  
\begin{figure*}[htbp]
\begin{center}
\includegraphics[width=2\columnwidth]{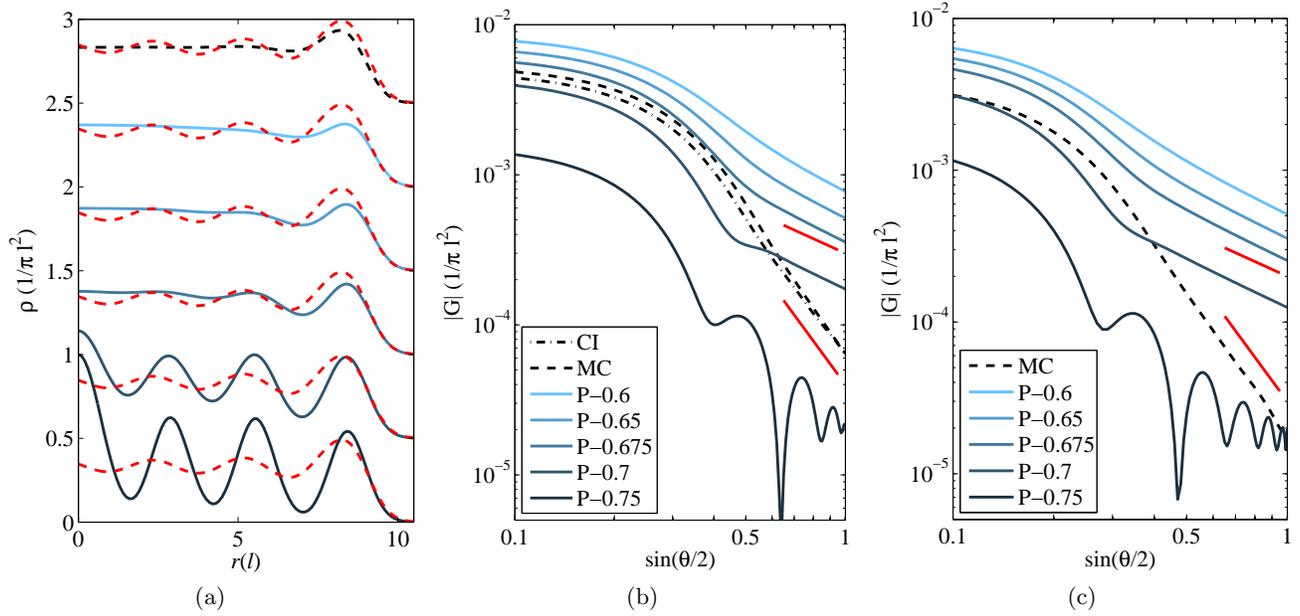}
\caption{(Color online) (a) 30 electron density profiles corresponding to Fig.~\ref{moc}(c) shifted vertically by 0.5 unit. Red dashed line is the phenomenological
estimate for the density oscillations at the thermodynamic limit. (b),(c) The decay of the edge Green's function 
calculated at $r=\sqrt{3(N-1)}+1$ from the occupations in Fig.~\ref{moc}(a) and (b), respectively. The short red lines illustrate slopes $-3$ 
(theoretical prediction for $\nu=1/3$) and $-1$ (Fermi liquid).}
\label{mocc}
\end{center}
\end{figure*}

The occupation numbers obtained in such way for $N=10$, $N=20$, and $N=30$ are presented in Fig.~\ref{moc}. 
As seen in the exact result (first row in 
Fig.~\ref{moc}(a)) the long-range Coulomb correlations cause oscillations in the occupations around $1/3$ (CI),
apart from the edge density modulation not present in the Laughlin's occupation numbers (MC). Sliding $\alpha$ from 0.6 to 0.75 gradually
strengthens these oscillations. P-0.65 is close to the Laughlin's occupations while P-0.675 is close
to the exact occupations (CI). Similar behavior is seen at larger particle numbers in Figs. \ref{moc} (b) and (c),
where P-0.675 yields again occupations plausibly closest to the exact unknown result.

The oscillations in the occupation numbers reflect the formation of an edge striped phase. An extrapolated phenomenological formula
for the slow-decaying charge density oscillations at the $\nu=1/3$ edge at the thermodynamic limit is given in Ref. \onlinecite{tsiper}
\begin{equation}
\rho(s)=\frac{1}{6}(\textrm{Erf}(s)+1)\left[1+\frac{1}{2}J_0\left(\tfrac{\pi}{2}(s-1)\right)\right]\  
\label{rhos}
\end{equation}
where $s/\sqrt{2}$ is the distance from the edge located at $\sqrt{3(N-1)}$, Erf is the Gauss error function, and $J_0$
is the Bessel function of the  first kind. Fig.~\ref{mocc}(a) shows the radial charge densities $\langle\Psi|\psi^{\dagger}(r)\psi(r)|\Psi\rangle$ calculated from the 30 electron
occupation numbers compared to the extrapolated formula (red dashed line). The latter has slightly longer oscillation wavelength 
compared to the P-$\alpha$ results while the amplitude of oscillations suggests that the optimal value of $\alpha$ is
somewhere between 0.675 and 0.7. Thus, although Eq.~(\ref{rhos}) has zero free parameters, it matches the 1-RDM results reasonably well.

Edge Green's function $G_{\rm E}$ is the amplitude for electron to propagate a distance along the edge. In the quantum Hall droplet,
the distance is related to the angle $\theta$ between the two points, and 
$G_{\rm E}=\langle\Psi|\psi^{\dagger}(z_0e^{i\theta})\psi(z_0)|\Psi\rangle$, where $z_0$ is a point
of the edge. Chiral Luttinger liquid theory of the fractional quantum Hall edge predicts the universal asymptotic behavior\cite{wen1,wen2}
\begin{equation}
|G_{\rm E}|\propto |z_0e^{i\theta}-z_0|^{-g}\propto|\sin(\theta/2)|^{-g}\ ,
\end{equation}
where $g=3$ for $\nu=1/3$. Values of $g\neq1$ lead to non-Ohmic current-voltage dependence $I\propto V^g$ in the tunneling
experiments. However, thus observed experimental value, $g\approx2.2-2.8$, is contrary to the theory possibly sample dependent.\cite{chang,grayson}

The decay of $|G_{\rm E}|$ calculated with the density-matrix power functionals is compared to 
the exact (CI) and the Laughlin wave function's result (MC) in Figs.~\ref{mocc}(b) and (c) for $N=10$ and $N=20$, respectively 
(not shown case $N=30$ looks similar). The short lines signify power-law exponents $g=1$ and $g=3$. Except for the
curve corresponding to P-0.75, which oscillates heavily, the P-$\alpha$ curves follow closely 
the  theoretical dashed black line until $\sin(\theta/2)\approx0.3$, after which they sheer off the course to yield an exponent 1.
This is due to the incorrect weights of the occupation numbers near the edge of the system and
investigated further in the next subsection where we apply 2-RDMFT to a smaller system.

The interaction energies are shown in Table \ref{tabe}. The exactness of the 
Laughlin trial wave function's energy (MC) up to 0.1\% for $N=10$ is expected to carry on to the larger electron numbers.
The interaction energy is seen to increase as a function of $\alpha$ in P-$\alpha$ and is optimal with
the functional P-0.7, which attains 99.9\% of the interaction energy for $N=20$ and $30$. 
Typical to 1-RDMFT calculations in general, the energies are mostly
below the assumed nearly exact Monte-Carlo energy.
\begin{table}[b]
\caption{Interaction energy in units of $e^2/\epsilon l$ at angular momentum $M=3N(N-1)/2$
and the percentage captured of the energy of the Laughlin wave function for density-matrix power functionals P-$\alpha$. }
\label{tabe}
\begin{tabular}{l|r|r|r|}
            & $N=10$& $N=20$& $N=30$\\
\hline
CI          &10.14&     &     \\
MC          &10.15&32.92&64.00\\
P-0.6       & 8.93&30.36&60.08\\
P-0.65      & 9.58&31.72&62.15\\
P-0.675     & 9.87&32.34&63.10\\
P-0.7       &10.12&32.88&63.93\\
P-0.75      &10.44&33.44&64.80\\
\end{tabular}
\begin{tabular}{|r|r|r|}
$N=10$& $N=20$& $N=30$\\
\hline
99.9\% &&      \\
100.0\%&100.0\%&100.0\%\\
88.0\% & 92.2\%& 93.9\%\\
94.4\% & 96.4\%& 97.1\%\\
97.2\% & 98.2\%& 98.6\%\\
99.7\% & 99.9\%& 99.9\%\\
102.9\%&101.6\%&101.2\%\\
\end{tabular}
\caption{Interaction energy of one elementary quasihole excitation in units of $e^2/\epsilon l$  and the percentage captured
of the interaction energy of the wave functional quasihole model (MC).}
\label{tabe3}
\begin{tabular}{l|r|r|r|}
            & $N=10$& $N=20$& $N=30$\\
\hline
MC          &-0.314&-0.497&-0.635\\
P-0.6       &-0.396&-0.601&-0.758\\
P-0.65      &-0.381&-0.588&-0.736\\
P-0.675     &-0.376&-0.583&-0.733\\
P-0.7       &-0.376&-0.584&-0.689\\
P-0.75      &-0.378&-0.585&-0.326\\
\end{tabular}
\begin{tabular}{|r|r|r|}
$N=10$& $N=20$& $N=30$\\
\hline
100\%&100\%&100\%\\
126\%&121\%&119\%\\
121\%&118\%&116\%\\
120\%&117\%&115\%\\
120\%&118\%&109\%\\
120\%&118\%&51\%\\
\end{tabular}
\end{table} 

While knowledge of the ground state energy may be useful when comparing different methods, only energy
differences are physically meaningful. In the RDM methods, we can calculate the energy differences between
lowest energy states of different $(M,N)$ sectors such as addition energy (change in $N$) as well as quasiparticle and some edge
excitations ($M$ changes). If the excited state becomes the ground state for some parameters, the Gilbert's 
theorem guarantees the existence of a 1-RDM functional minimized by the exact 1-RDM but even if this is not
the case, a good functional might still exist.

As mentioned previously, some of the cusps in $M$-$V_{\rm ee}$-curves correspond to the quasiparticle excitations
of stable quantum Hall phases. Next, we consider such a quasihole excitation above the $\nu=1/3$ state.
The quasihole is a charged vortex carrying fractional charge $q=e/3$ and obeying anyonic statistics.\cite{arovas,camino}
To obtain interaction part of the quasihole excitation energy at $\nu=1/3$, we need to calculate the difference
 $V_{\rm ee}(M_{1/3}+N,N)-V_{\rm ee}(M_{1/3},N)$. The angular momentum
$M_{1/3}+N$ follows from the Laughlin's quasihole wave function, which is also used to compute an estimate for 
$V_{\rm ee}(M_{1/3}+N,N)$ with Monte Carlo. Due to the fact that Laughlin's wave function is more accurate than
the quasihole wave function, variational principle implies that the Monte Carlo estimate  to the (negative) contribution 
to the excitation energy is likely an upper bound to the exact result. Nevertheless,
for 8 particles the difference to exact CI result is less than 0.2\% so we expect the estimates to be quite accurate.
 The quasihole wave function reads
\begin{equation}
\Psi^{1/3}_{\textrm{L}}(\{z_i\})=\prod_i(z_i-z_{\textrm{CM}})\prod_{i<j}(z_i-z_j)^3e^{-\frac{1}{2}\sum_i z_i\bar{z}_i}\ ,
\end{equation}
where $z_{\textrm{CM}}=(1/N)\sum_i z_i$. The interaction contribution to the excitation energy is shown in
Table \ref{tabe3}. The power functionals appear to overestimate the energy gap by 10 to 20 percent compared to the trial
wave function though the results seem to get more accurate with increasing $N$. This preliminary result indicates that the method could 
prove useful in assessing the stability of different models for quantum Hall phases characterized by certain angular momentum and spin.
Additionally, 1-RDM method offers a simple framework to include the higher Landau levels, however, instead of the bare eigenvalues of the
reduced density matrix, one must then optimize the eigenvectors also.

\subsection{2-RDMFT results for three electrons}
\label{res3}

In this final part, we apply the exact 2-RDM functional (Eq.~(\ref{exa})) to a three electron droplet
in the 1/3 state again with the maximum single-particle angular momentum set to $3(N-1)$. We will see that the resulting 2-RDM, though not strictly physical, is close to the exact solution 
and yields better results than our 1-RDM functionals.

Compared to the 1-RDM calculations seen above, the computational cost of the problem in the 2-RDM optimization is considerably 
larger and scales at higher order $p^6$ (versus $p^4$) with the number of single-particle orbitals $p$. The 1/3 Laughlin state 
for 3 electrons has angular momentum $M=9$ and requires only 7 single-particle states. However, the number of
optimization variables in $\Gamma$, $Q$, and $G$ are 276, 378, and 1225, respectively, and the constraint equations (\ref{ntr},\ref{qg},\ref{mc1},\ref{mc2}) 
together lead to 2798 constraints each facilitated by a Lagrange multiplier. In practice, this means that
this method takes more time than exact diagonalization in any system that could be solved in a reasonable time.
However, owing to the  exponential scaling of the exact diagonalization problem, the situation could change in future 
with development of faster computers and more efficient semi-definite programming and optimization algorithms.

\begin{figure}[b]
\begin{center}
\includegraphics[height=0.9\columnwidth]{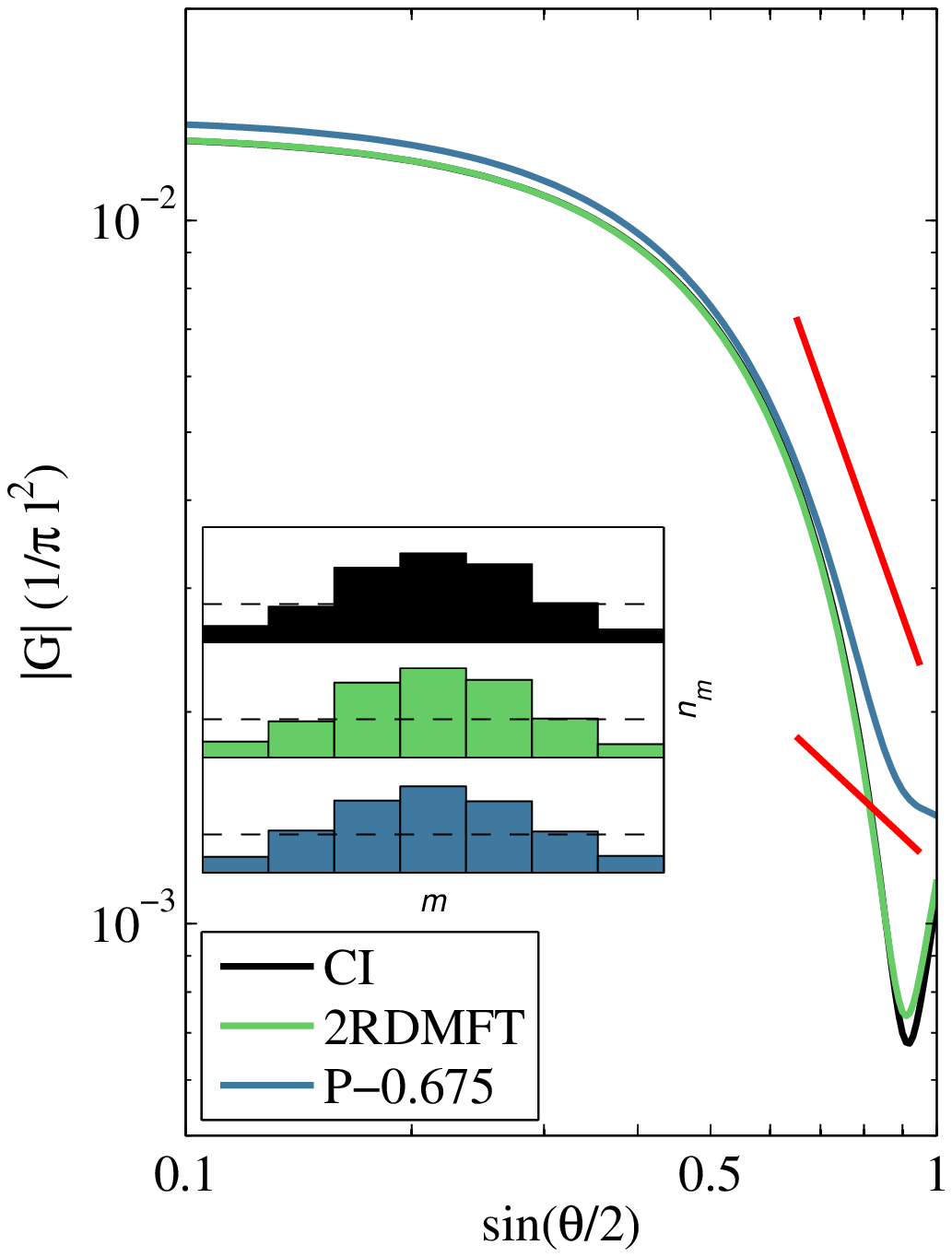}\\
\caption{(Color online) The decay of the edge Green's function corresponding to the inset 3-electron 
occupation numbers. As previously, the short red lines illustrate slopes $-g=-3$ and $-1$, and $G_{\rm E}$
is evaluated at $r=\sqrt{3(N-1)}+1$.}
\label{3g}
\end{center}
\begin{center}
\includegraphics[width=1.0\columnwidth]{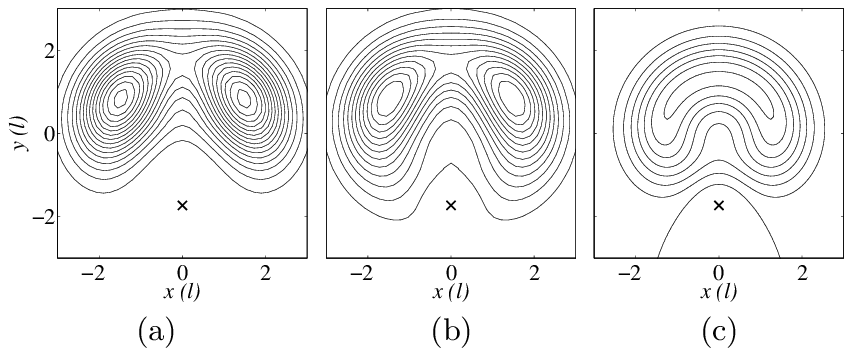}\\
\caption{The pair-correlation function $g(z_1,z_2)=\rho_2(z_1,z_2)/\rho(z_1)$ with
the first coordinate placed at the density maximum of the negative y-axis $z_1=-i\sqrt{3}$ for (a) exact diagonalization, (b) 2-RDMFT,
and (c) 1-RDMFT with P-0.675. Contours are separated by 0.05 $1/\pi l^2$ and start from 0.05 $1/\pi l^2$ in (a) and (b) and from 0 in (c).}
\label{pdds}
\end{center}
\end{figure} 

\begin{figure*}[t!]
\begin{center}
\includegraphics[width=2.0\columnwidth]{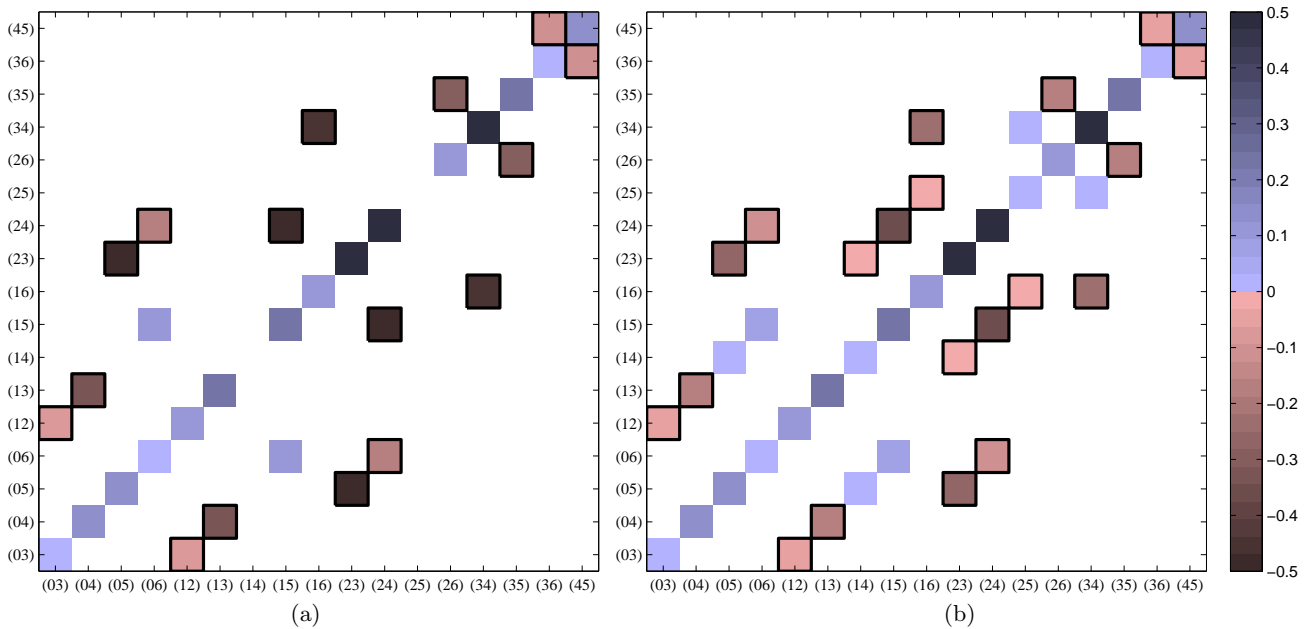}\\
\caption{(Color online) Illustration of the 2-RDM $\Gamma^{(ij)}_{(kl)}$ for three electrons in the $\nu=1/3$ state computed with
(a) exact diagonalization and (b) 2-RDMFT. The same color bar applies to both figures while zero values are left white. The
negative matrix elements are indicated by black edge.}
\label{gammaob}
\end{center}
\end{figure*}

Figure \ref{3g} shows the occupation numbers and the decay of the edge Green's function for three electrons at $\nu=1/3$ ($M=3N(N-1)/2=9$).
Due to a finite size effect, the exact diagonalization Green's function has a downward cusp at $\sin(\theta/2)\approx0.9$.
The 2-RDMFT result has a similar cusp while the P-0.675 result, which turns smoothly to a lower slope decay, does not have one.
We verified that this is due to the difference in the weights of the last three occupation numbers corresponding to the edge of the system.
Since the HF solution would have exponent $g=1$, the correct decay property of the Green's function follows from the off-diagonal terms
of the interaction operator. The 1-RDMFT that only uses the diagonal $V_{ijij}$ and $V_{ijji}$ terms of the interaction matrix can not yield
the correct behavior unless we have a very good density matrix functional.

Recall that the backbone of the 1-RDMFT was the approximation of the pair-density.
Figure \ref{pdds} shows the pair-correlation functions corresponding again to exact diagonalization, 2-RDMFT, and P-0.675, where the
latter is reconstructed from the 1-RDM using Eq.~(\ref{apd}). Although the 1-RDM reconstructed pair-correlation function is reasonable
vanishing at the position of the fixed electron, only the 2-RDMFT is able to produce the two-peak structure of the exact 
result with reduced density along the $y$-axis.
 
Granted that the interaction energy functional in the 2-RDMFT is exact, the results still do not coincide with the exact diagonalization 
results because the 3-representability conditions that ensure the physicality in this 3-electron system can not be taken into account
without invoking the generalization of 2-RDMFT to include higher order RDMs. Figure~\ref{gammaob}(a) shows the exact non-zero matrix elements of
the 2-RDM $\Gamma^{(ij)}_{(kl)}$ in the antisymmetric basis $|(ij)\rangle=(|ij\rangle-|ji\rangle)/2$ and (b) calculated with the 2-RDMFT. 
The largest discrepancy between the results is the vanishing of the matrix elements involving states $|(14)\rangle$ or $|(25)\rangle$ for the exact result, while
for example $\Gamma^{(05)}_{(14)}\neq0$ in the 2-RDMFT result. The matrix elements should vanish, as they are related to the fictitious
many-body basis states that have double occupancy of orbital 4 or 2 ($1+4+4=2+2+5=9$, the total angular momentum). Consequently, the absolute values
of the matrix elements also differ. However, the computed pair-correlation function and edge Green's function suggest that many of the physical quantities are
not significantly affected by the absence of exact $N$-representability, and that the lack of computer power might be the only real stumbling stone in the way of the 2-RDMFT.

\section{Conclusions}
\label{summary}
In summary, we have applied the one-body reduced density-matrix functional theory to small and large quantum
Hall droplets at the spin-polarized strong magnetic field regime. The density-matrix power functional seems to work reasonably well at the 
strongly correlated $\nu\ll1$ regime where the occupation numbers of the natural orbitals are small. The newly applied method yields 
previously inaccessible valuable information with large particle numbers about the energetics and quantities that derive
from the one-body reduced density matrix. The density-matrix power functional yields reasonable bulk densities with the power parameter in the range
0.65-0.7.
However, the detailed properties of the edge are not produced accurately with this functional.
Moreover, it is not known if a good functional for a specific quantum Hall state would work universally at different filling fractions.
Nevertheless, the computationally expensive 2-body reduced density matrix method seems to facilitate the properties of the edge, though this should
be verified with a larger electron number in future.

Prospects of the 1-RDMFT in quantum Hall systems include generalizations to spin and multiple Landau levels. 
Studies with systems without edge (sphere) 
and non-trivial topology (torus) are also encouraged while new state of the art energy functionals are of course very welcome.

\section*{Acknowledgements}

This study has been supported by the Academy of Finland through its Centres of Excellence Program (2006-2011). 
ET acknowledges financial support from the Vilho, Yrj\"o, and Kalle V\"ais\"al\"a Foundation of the Finnish
Academy of Science and Letters. We also thank I. Makkonen for careful reading of the manuscript and E. R\"as\"anen and 
R. van Leeuwen for useful discussions.

\end{document}